\documentclass[showpacs,aps,graphicx,twocolumn]{revtex4}
\usepackage{graphicx}
\begin{document}

\title{Optimal entanglement concentration for quantum dot and optical microcavities systems}

\author{ Yu-Bo Sheng,$^{1,2}$\footnote{Email address:
shengyb@njupt.edu.cn} Lan Zhou,$^{2,3}$ Lei Wang,$^{1,2}$ and
Sheng-Mei Zhao,$^{1,2}$}
\address{$^1$ Institute of Signal Processing  Transmission, Nanjing
University of Posts and Telecommunications, Nanjing, 210003,  China\\
$^2$ Key Lab of Broadband Wireless Communication and Sensor Network
 Technology, Nanjing University of Posts and Telecommunications, Ministry of
 Education, Nanjing, 210003,
 China\\
$^3$College of Mathematics \& Physics, Nanjing University of Posts and Telecommunications, Nanjing,
210003, China\\}

\date{\today}

\begin{abstract}
A recent paper [Chuan Wang, Phys. Rev. A \textbf{86}, 012323 (2012)] discussed an entanglement concentration protocol (ECP)
for partially entangled electrons using a quantum dot and microcavity coupled system. In his paper,  each two-electron spin system in a partially entangled state can be concentrated with the assistance of an ancillary quantum dot and a single photon. In this paper, we will present
an optimal ECP for such entangled electrons with the help of only one single photon. Compared with the protocol of Wang, the most significant advantage is that during the whole ECP, the single photon only needs to pass through one microcavity which will  increase the total success probability if the
 cavity is imperfect. The whole protocol can be repeated to get a higher success probability.
With the feasible technology, this protocol may be useful in current long-distance quantum communications.
\end{abstract}

\pacs{ 03.67.Bg, 42.50.Pq, 78.67.Hc, 78.20.Ek} \maketitle
\section{Introduction}
Entanglement plays an important role in current quantum information
processing \cite{book,rmp}. For most of the practical
quantum
 communication and computation protocols, the maximally entangled
states are usually required. For example, quantum key distribution
\cite{rmp,Ekert91}, quantum teleportation
\cite{teleportation}, quantum secret sharing
\cite{QSS1,QSS2,QSS3}, quantum secure direction communication
\cite{QSDC1,QSDC2,QSDC3} and  quantum dense coding \cite{densecoding} all need
entanglement to set up the quantum channels. However, the
entanglement channel will inevitably decrease because it always
contacts the environment. The degraded entanglement will decrease
the fidelity of the teleportation, or make some quantum
communication protocols insecure. Therefore, people should seek for
effective ways to combat noise and recover the entanglement to a
high quality.

Entanglement concentration
\cite{C.H.Bennett2,swapping1,swapping2,zhao1,Yamamoto1,shengpra2,shengqic,shengpra3,shengwstateconcentration,wangchuan,wangchuan2,dengpra,wanghf}
is one of the powerful methods which is used to improve the quality of
the entanglement. It can distill a subset system in a maximally entanglement state from
a set of systems in a partially entangled (less-entangled) pure
state. In 1996, Bennett \emph{et al.} proposed an entanglement
concentration protocol (ECP) based on the Schmidt decomposition
\cite{C.H.Bennett2}. In the protocol, some collective measurements are needed, which
are hard to manipulate in experiment at present. Bose \emph{et al.}
proposed an ECP based on entanglement swapping \cite{swapping1}.
Later, this method was developed by Shi \emph{et al.} with collective
unitary evaluation \cite{swapping2}. Zhao\emph{ et al.} and Yamamoto
\emph{et al.} proposed two similar ECPs with linear optics
independently \cite{zhao1,Yamamoto1}. In 2008, ECPs based on
the cross-Kerr nonlinearity was proposed \cite{shengpra2}.

Currently, most of the  ECPs are focused on photons, for photons are the
best candidate for optical transmission. Actually, quantum
communication and computation can also be achieved with solid
electrons \cite{beenakker,waks,cavity,wangrepeater,hu1,hu2,hu3,hu4,xu,wangchuan3,litao}. For example, In 2004, Beenakker \emph{et al.} showed that with the
help of charge degree of freedom, they could brake through the obstacle of
the no-go theorem and construct a CNOT gate
\cite{beenakker}. Moreover, Waks and Vuckovic discussed the
interaction of a cavity with a dipole \cite{waks}. The cavity decay
rate is larger than the vacuum Rabi frequency. It has been used to
construct a quantum repeaters in a weak-coupling regime
\cite{cavity,wangrepeater}.
Wang \emph{et al.} also proposed an ECP with electron-spin entangled
states using quantum dot spins in optical microcavities
\cite{wangchuan}. Recently, he improved the protocol, and presented
an efficient ECP with the help of an ancillary quantum dot and a single photon \cite{wangchuan2}. However, this protocol is still not an optimal one.

In this paper, we present an optimal ECP for electronic systems
by exploiting a weak-coupling regime. In this protocol, only one
pair of less-entangled pure state and a single photon are required.
Compared with the conventional ECPs, this ECP resorts less original
less-entangled pure state sources. It can also  reach the same
success probability as  described in Ref. \cite{wangchuan}.
Moreover, it can be repeated to get a higher success
probability. Compared with Ref. \cite{wangchuan2}, we do not require
the single quantum dot as an ancillary and only the single photon can complete the task.
Moreover, the single photon only needs to pass through one  microcavity, which will greatly
improve the success probability in a practical situation.

This paper is organized as follows. In Sec. \textbf{2}, we first briefly
explain the basic element of this protocol, which is also shown in
Ref. \cite{wangchuan,cavity}. In Sec. \textbf{3}, an ECP assisted with
single photon is described. In
Sec. \textbf{4},  we present a discussion and  summary.

\section{Basic element for ECP}
Before we start to explain our ECP, we first introduce the basic
element of our protocol, as shown in Fig. 1. In Ref. \cite{cavity}, it
can be used to perform the CNOT gate and the Bell-state analysis. It
also has been discussed to implement the photon entangler,
entanglement beam splitter, optical  Faraday rotation
\cite{hu1,hu2,hu3}. Recently, Hu and Rarity also presented schemes
for efficient state teleportation and entanglement swapping, using
single quantum dot spin in the optical microcavity \cite{hu4}. The
single-electron-charged quantum dot in a resonator shows a good
interaction between a photon and an electron spin. The photon
and the electron can be used to generate the hybrid entanglement
with the quantum dot coupled to a microcavity.
From Fig. 1, if we consider the spin of the electron in spin up
state $|\uparrow\rangle$ and a photon in state $s_{z}=+1$, the
circularly polarized light might change their polarization according
to direction of propagation, and the spin of the electron, after the
photon passing through the cavity. For example, if the propagation
of the photon is in the direction of the z axis, and the polarization
of the photon is  right-circular-polarization, say
$|R^{\uparrow}\rangle$, it will become $|L^{\downarrow}\rangle$, if
the electron is $|\uparrow\rangle$. The total rules of the state
change under the interaction of the photon with $s_{z}=\pm1$ can be
described as \cite{wangchuan,cavity}
\begin{eqnarray}
&&|R^{\uparrow},\uparrow\rangle\rightarrow|L^{\downarrow},\uparrow\rangle,
\hspace{8mm}|R^{\downarrow},\uparrow\rangle\rightarrow
-|R^{\downarrow},\uparrow\rangle,\nonumber\\
&&|R^{\uparrow},\downarrow\rangle\rightarrow
-|R^{\uparrow},\downarrow\rangle,
\hspace{5mm}|R^{\downarrow},\downarrow\rangle\rightarrow
|L^{\uparrow},\downarrow\rangle,\nonumber\\
&&|L^{\uparrow},\uparrow\rangle\rightarrow
-|L^{\uparrow},\uparrow\rangle,\hspace{5.5mm}|L^{\downarrow},
\uparrow\rangle\rightarrow|R^{\uparrow},\uparrow\rangle,\nonumber\\
&&|L^{\uparrow},\downarrow\rangle\rightarrow|R^{\downarrow},\downarrow\rangle,\hspace{8mm}
|L^{\downarrow},\downarrow\rangle\rightarrow
-|L^{\downarrow},\downarrow\rangle.
\end{eqnarray}
Here $|R\rangle$ and $|L\rangle$ denote the states of
right-circular-polarized and left-circular-polarized photons,
respectively. The $\uparrow$ and $\downarrow$ on the superscript
arrow are the propagation direction along the z axis.

\begin{figure}[!h]
\begin{center}
\includegraphics[width=3cm,angle=0]{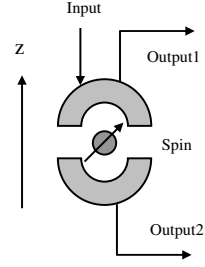}
\caption{A schematic drawing of the
basic element of our ECP. The quantum dot spin is coupled in optical
microcavities.  Input represents the input port of a photon. Output1
and Output2 are the output ports of the photon after coupled with
the electron-spin system.}
\end{center}
\end{figure}

\section{ECP  assisted with only a single photon}
In this section, we will show that the single
electron is not necessary, only a single photon can also complete this task, which leads it more optimal than the protocols in Refs. \cite{wangchuan,wangchuan2}.

Suppose Alice and Bob share the less-entangled state of the form
\begin{eqnarray}
|\phi^{+}\rangle_{12}=\alpha|\uparrow\rangle_{1}|\uparrow\rangle_{2}+\beta|\downarrow\rangle_{1}|\downarrow\rangle_{2},\label{less1}
\end{eqnarray}
with $|\alpha|^{2}+|\beta|^{2}=1$. The subscripts "1" and "2" are
the spin 1 and spin 2 shown in Fig. 2. Alice prepares another single photon as
\begin{eqnarray}
|\Phi\rangle_{P}=\alpha|R\rangle+\beta|L\rangle.\label{singlephoton}
\end{eqnarray}

\begin{figure}[!h]
\begin{center}
\includegraphics[width=7cm,angle=0]{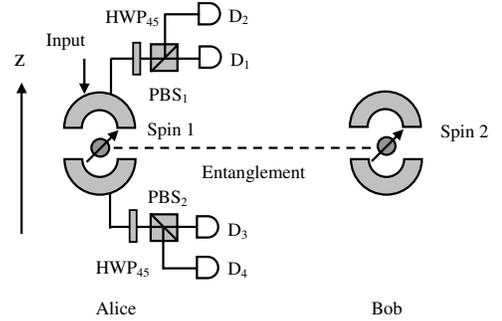}
\caption{A schematic drawing of the
basic principle of our ECP. The PBSs are the polarization beam splitters. They are used
to transmit the photon in $|H\rangle$ polarization  and
reflect the photon in $|V\rangle$ polarization. HWP$_{45}$ is the quarter wave plate.}
\end{center}
\end{figure}

The less-entangled state combined with the single photon evolves as
\begin{eqnarray}
&&|\Phi\rangle_{P}|\phi^{+}\rangle_{12}=(\alpha|R^{\downarrow}\rangle+\beta|L^{\downarrow}\rangle)\nonumber\\
&&(\alpha|\uparrow\rangle_{1}|\uparrow\rangle_{2}+\beta|\downarrow\rangle_{1}|\downarrow\rangle_{2})\nonumber\\
&=&\alpha^{2}|R^{\downarrow}\rangle|\uparrow\rangle_{1}|\uparrow\rangle_{2}+\beta^{2}|L^{\downarrow}\rangle|\downarrow\rangle_{1}|\downarrow\rangle_{2}\nonumber\\
&+&\alpha\beta(|R^{\downarrow}\rangle|\downarrow\rangle_{1}|\downarrow\rangle_{2}+|L^{\downarrow}\rangle|\uparrow\rangle_{1}|\uparrow\rangle_{2})\nonumber\\
&\rightarrow&-\alpha^{2}|R^{\downarrow}\rangle|\uparrow\rangle_{1}|\uparrow\rangle_{2}-\beta^{2}|L^{\downarrow}\rangle|\downarrow\rangle_{1}|\downarrow\rangle_{2}\nonumber\\
&+&\alpha\beta(|L^{\uparrow}\rangle|\downarrow\rangle_{1}|\downarrow\rangle_{2}+|R^{\uparrow}\rangle|\uparrow\rangle_{1}|\uparrow\rangle_{2}).\nonumber\\\label{combine3}
\end{eqnarray}
Alice then lets her photon pass through the quarter wave plate (HWP$_{45}$), which makes
$|H\rangle=\frac{1}{\sqrt{2}}(|R\rangle+|L\rangle)$, and
$|V\rangle=\frac{1}{\sqrt{2}}(|R\rangle-|L\rangle)$. $|H\rangle$ and $|V\rangle$ represent the horizonal and
vertical polarization, respectively.
Obviously, the items
$-\alpha^{2}|R^{\downarrow}\rangle|\uparrow\rangle_{1}|\uparrow\rangle_{2}$
and
$-\beta^{2}|L^{\downarrow}\rangle|\downarrow\rangle_{1}|\downarrow\rangle_{2}$
will make the detectors D$_{3}$ or D$_{4}$ fire, while the items
$|L^{\uparrow}\rangle|\downarrow\rangle_{1}|\downarrow\rangle_{2}$
and $|R^{\uparrow}\rangle|\uparrow\rangle_{1}|\uparrow\rangle_{2}$
will make the conventional single-photon detectors D$_{1}$ or D$_{2}$ fire. Therefore, after
passing through the HWP$_{45}$, the state
$\frac{1}{\sqrt{2}}(|L^{\uparrow}\rangle|\downarrow\rangle_{1}|\downarrow\rangle_{2}+|R^{\uparrow}\rangle|\uparrow\rangle_{1}|\uparrow\rangle_{2})$
will  become
\begin{eqnarray}
&&\frac{1}{\sqrt{2}}(|L^{\uparrow}\rangle|\downarrow\rangle_{1}|\downarrow\rangle_{2}+|R^{\uparrow}\rangle|\uparrow\rangle_{1}|\uparrow\rangle_{2})\nonumber\\
&\rightarrow&\frac{1}{2}(|H\rangle-|V\rangle)|\downarrow\rangle_{1}|\downarrow\rangle_{2}
+\frac{1}{2}(|H\rangle+|V\rangle)|\uparrow\rangle_{1}|\uparrow\rangle_{2}\nonumber\\
&=&\frac{1}{\sqrt{2}}|H\rangle\frac{1}{\sqrt{2}}(|\uparrow\rangle_{1}|\uparrow\rangle_{2}+|\downarrow\rangle_{1}|\downarrow\rangle_{2})\nonumber\\
&+&\frac{1}{\sqrt{2}}|V\rangle\frac{1}{\sqrt{2}}(|\uparrow\rangle_{1}|\uparrow\rangle_{2}-|\downarrow\rangle_{1}|\downarrow\rangle_{2}).
\end{eqnarray}
Finally, after passing through the PBS$_{1}$, which transmits the
$|H\rangle$ polarization photon and reflects the
$|V\rangle$ polarization photon, they will obtain
$\frac{1}{\sqrt{2}}(|\uparrow\rangle_{1}|\uparrow\rangle_{2}+|\downarrow\rangle_{1}|\downarrow\rangle_{2})$
if D$_{1}$ fires, and obtain
$\frac{1}{\sqrt{2}}(|\uparrow\rangle_{1}|\uparrow\rangle_{2}-|\downarrow\rangle_{1}|\downarrow\rangle_{2})$,
if D$_{2}$ fires. The success probability is
$2|\alpha\beta|^{2}$.

Interestingly, from Eq. (\ref{combine3}), it has another case that
the photon will be in another output mode, which makes the
original state collapse to
$\alpha^{2}|R^{\downarrow}\rangle|\uparrow\rangle_{1}|\uparrow\rangle_{2}+\beta^{2}|L^{\downarrow}\rangle|\downarrow\rangle_{1}|\downarrow\rangle_{2}$.
After passing through the HWP$_{45}$ and PBS$_{2}$, if the detector
D$_{3}$ fires, they will obtain
$\alpha^{2}|\uparrow\rangle_{1}|\uparrow\rangle_{2}+\beta^{2}|\downarrow\rangle_{1}|\downarrow\rangle_{2}$,
and if the D$_{4}$ fires, they will obtain
$\alpha^{2}|\uparrow\rangle_{1}|\uparrow\rangle_{2}-\beta^{2}|\downarrow\rangle_{1}|\downarrow\rangle_{2}$.
Both of them are the less-entangled states, and can be reconcentrated into a maximally entangled pair in the second concentration round.
Briefly speaking, if they get
\begin{eqnarray}
|\phi^{+}\rangle'_{12}=\alpha^{2}|\uparrow\rangle_{1}|\uparrow\rangle_{2}+\beta^{2}|\downarrow\rangle_{1}|\downarrow\rangle_{2}.\label{less2}
\end{eqnarray}
 Alice only needs to choose another single photon
of the form
\begin{eqnarray}
|\Phi\rangle'_{P}=\alpha^{2}|R\rangle+\beta^{2}|L\rangle.
\end{eqnarray}
So the whole system can be written as
\begin{eqnarray}
&&|\phi^{+}\rangle'_{12}|\Phi\rangle'_{P}\nonumber\\
&=&(\alpha^{2}|\uparrow\rangle_{1}|\uparrow\rangle_{2}+\beta^{2}|\downarrow\rangle_{1}|\downarrow\rangle_{2})(\alpha^{2}|R^{\downarrow}\rangle+\beta^{2}|L^{\downarrow}\rangle)\nonumber\\
&=&\alpha^{4}|\uparrow\rangle_{1}|\uparrow\rangle_{2}|R^{\downarrow}\rangle+\beta^{4}|\downarrow\rangle_{1}|\downarrow\rangle_{2}|L^{\downarrow}\rangle\nonumber\\
&+&\alpha^{2}\beta^{2}(|\uparrow\rangle_{1}|\uparrow\rangle_{2}|L^{\downarrow}\rangle+|\downarrow\rangle_{1}|\downarrow\rangle_{2}|R^{\downarrow}\rangle)\nonumber\\
&\rightarrow&-\alpha^{4}|R^{\downarrow}\rangle|\uparrow\rangle_{1}|\uparrow\rangle_{2}-\beta^{4}|L^{\downarrow}\rangle|\downarrow\rangle_{1}|\downarrow\rangle_{2}\nonumber\\
&+&\alpha^{2}\beta^{2}(|L^{\uparrow}\rangle|\downarrow\rangle_{1}|\downarrow\rangle_{2}+|R^{\uparrow}\rangle|\uparrow\rangle_{1}|\uparrow\rangle_{2}).\nonumber\\\label{combine4}
\end{eqnarray}
Obviously, from Eq. (\ref{combine4}), following the same principle described above,  the photon will pass through the
optical cavity and will be detected. If the detectors D$_{1}$
or D$_{2}$ fires, they will obtain the maximally entangled pair. The success probability is $
\frac{2|\alpha\beta|^{4}}{|\alpha|^{4}+|\beta|^{4}}$.
If the detectors D$_{3}$
or D$_{4}$ fires, they will obtain another less-entangled pair of the form
\begin{eqnarray}
|\phi^{\pm}\rangle''_{12}=\alpha^{4}|\uparrow\rangle_{1}|\uparrow\rangle_{2}\pm\beta^{4}|\downarrow\rangle_{1}|\downarrow\rangle_{2}.\label{less3}
\end{eqnarray}

It is still a less-entangled state
 which
can also be reconcentrated for the third round. In this way, by repeating the ECP, they can
obtain a higher success probability than other ECPs. The success probability
in each concentration round can be written as
\begin{eqnarray}
P_{1}&=&2|\alpha\beta|^{2}.\nonumber\\
P_{2}&=&\frac{2|\alpha\beta|^{4}}{|\alpha|^{4}+|\beta|^{4}}.\nonumber\\
P_{3}&=&\frac{2|\alpha\beta|^{8}}{(|\alpha|^{4}+|\beta|^{4})(|\alpha|^{8}+|\beta|^{8})},\nonumber\\
&\cdots\cdots&\nonumber\\
P_{K}&=&\frac{2|\alpha\beta|^{2^{K}}}{(|\alpha|^{4}+|\beta|^{4})(|\alpha|^{8}+|\beta|^{8})\cdots(|\alpha|^{2^{K}}+|\beta|^{2^{k}})}.\label{probability}
\end{eqnarray}

Actually, the realization of this ECP relies on the efficiency of transmission and  reflection for electrons and photon described in Sec. \textbf{2}.
We can
calculate the practical transmission and reflection coefficients,
according to
Heisenberg equations of motion for the cavity-field operator and the
trion dipole operator in weak excitation approximation.
 The reflection and
transmission coefficients can be written as
\begin{eqnarray}
r(w)&=&1+t(\omega),\nonumber\\
t(\omega)&=&\frac{-\kappa[i(\omega_{X^{-}}-\omega)+\frac{\gamma}{2}]}{[i(\omega_{X^{-}}-\omega)+\frac{\gamma}{2}][i(\omega_{c}-\omega)+\kappa+\frac{\kappa_{s}}{2}]+g^{2}},\nonumber\\
\end{eqnarray}
where $g$ represents the coupling constant. $\frac{\gamma}{2}$ is the
$X^{-} $ dipole decay rate. $\kappa$ and $\kappa_{s}/2$ are the
cavity field decay rate into the input and output modes and the leaky
rate, respectively \cite{hu2}.
 In
the approximation of weak excitation,
$\omega_{c}=\omega_{X^{-}}=\omega_{0}$, and $g=0$, we can get the
reflection and transmission coefficients as
\begin{eqnarray}
r_{0}(\omega)=\frac{i(\omega_{0}-\omega)+\frac{\kappa_{s}}{2}}{i(\omega_{0}-\omega)+\frac{\kappa_{s}}{2}+\kappa},\nonumber\\
t_{0}(\omega)=\frac{-\kappa}{i(\omega_{0}-\omega)+\frac{\kappa_{s}}{2}+\kappa}.
\end{eqnarray}
Here the $\omega_{0}$, $\omega_{c}$ and
$\omega_{X^{-}}$ are the frequencies of the input photon, cavity
mode, and the spin-dependent optical transition, respectively.
The transmission and reflection operators can be rewritten as
\begin{eqnarray}
\hat{t}(\omega)=t_{0}(\omega)(|R\rangle\langle R|\otimes
|\uparrow\rangle\langle\uparrow|+|L\rangle\langle L|\otimes
|\downarrow\rangle\langle\downarrow|)\nonumber\\
+t(\omega)(|R\rangle\langle R|\otimes
|\uparrow\rangle\langle\uparrow|+|L\rangle\langle L|\otimes
|\downarrow\rangle\langle\downarrow|),\nonumber\\
\hat{r}(\omega)=r_{0}(\omega)(|R\rangle\langle R|\otimes
|\uparrow\rangle\langle\uparrow|+|L\rangle\langle L|\otimes
|\downarrow\rangle\langle\downarrow|)\nonumber\\
+r(\omega)(|R\rangle\langle R|\otimes
|\uparrow\rangle\langle\uparrow|+|L\rangle\langle L|\otimes
|\downarrow\rangle\langle\downarrow|).
\end{eqnarray}

Therefore, we can recalculate the success probability in each concentration round as
\begin{eqnarray}
&&P'_{1}=\frac{|r(\omega)|}{\sqrt{|r_{0}(\omega)|^{2}+|r(\omega)|^{2}}}P_{1}.\nonumber\\
&&P'_{2}=\frac{|t_{0}(\omega)|}{\sqrt{|t_{0}(\omega)|^{2}+|t(\omega)|^{2}}}\frac{|r(\omega)|}{\sqrt{|r_{0}(\omega)|^{2}+|r(\omega)|^{2}}}P_{2}.\nonumber\\
&&P'_{3}=(\frac{|t_{0}(\omega)|}{\sqrt{|t_{0}(\omega)|^{2}+|t(\omega)|^{2}}})^{2}\frac{|r(\omega)|}{\sqrt{|r_{0}(\omega)|^{2}+|r(\omega)|^{2}}}P_{3}.\nonumber\\
&&\cdots\cdots\nonumber\\
&&P'_{K}=(\frac{|t_{0}(\omega)|}{\sqrt{|t_{0}(\omega)|^{2}+|t(\omega)|^{2}}})^{K-1}
\frac{|r(\omega)|}{\sqrt{|r_{0}(\omega)|^{2}+|r(\omega)|^{2}}}P_{K}.\nonumber\\\label{probability1}
\end{eqnarray}
The total success probability can be written as
\begin{eqnarray}
P_{t}=P'_{1}+P'_{2}+\cdots=\sum^{\infty}_{K=1}P'_{K}.
\end{eqnarray}
\begin{figure}[!h]
\begin{center}
\includegraphics[width=8cm,angle=0]{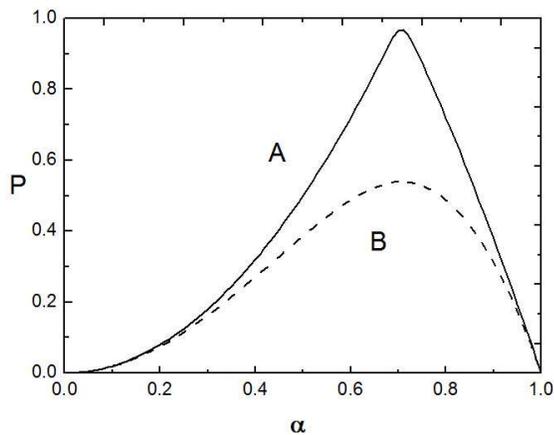}
\caption{Success probability $P$ for obtaining a maximally entangled
state after performing this ECP is altered with the initial
coefficient $\alpha\in(0,1)$.   Curve A is the ideal case with no
leakage. Curve B is the success probability with
 $\kappa_{s}=0.5\kappa$, $g=0.5\kappa$
and $\gamma=0.1\kappa$.  For numerical
simulation, we let $K=5$ as a good approximation.}
\end{center}
\end{figure}

We calculate the total success probability in both the ideal case with no leakage and with
 $\kappa_{s}=0.5\kappa$, $g=0.5\kappa$
and $\gamma=0.1\kappa$. In Fig. 3, it is shown that in the ideal case, the success probability
can reach the maximally value 1 when $\alpha=\frac{1}{\sqrt{2}}$. However, the leakage of the cavity
will decrease the success probability. We show that the maximum value of $P$ is about 0.5 when $\kappa_{s}=0.5\kappa$. For numerical
simulation, we let $K=5$ as a good approximation.

\section{Discussion and summary}
So far, we have fully explained our ECP.  In our ECP, we exploit a
single charged quantum dot inside an optical cavity. This single
charged quantum dot can be implemented by GaAs/InAs interface
quantum dot. Therefore, we require the long coherent time of the
quantum dot and the strong coupling of the quantum dot with the
cavity to ensure the photon can be fully coupled with quantum dot.
Fortunately, current experiment showed that the coherence time is
long enough of the GaAs- or InAs-based quantum dots \cite{xu}.
Moreover, current experiments also showed that the strong coupling
has also been observed in different systems
\cite{couple1,couple2,couple3}.

It is interesting to compare this ECP with other ECPs. In the Ref.
\cite{wangchuan}, Wang \emph{et al.} also proposed an ECP based on
quantum dot spins. In their protocol, in each step, they resort two
copies of less-entangled pairs. After measuring the photon, at least
one pair of less-entangled state should be destroyed, or both pairs
should be discarded. In the Ref. \cite{wangchuan2}, they require
a single charge qubit and a single photon to reach the same success probability
than the first one.
In our  protocol, we require one pair of
less-entangled state and a single photon to
reach the same success probability as Refs. \cite{wangchuan,wangchuan2}. In a practical conditional, this protocol
is more powerful. From Eq. (\ref{probability1}), the
total success probability is related with the efficiency of the transmission and reflection and the
photon will be lost if the cavity is imperfect.
 In this protocol, in each concentration round, the single photon only need
to pass through one  microcavity while in Ref. \cite{wangchuan2}, it should pass through two  microcavies.
So it will  decrease the total success probability if the cavity leakage is large.
Moreover, in his protocol, he should first prepare the assisted single
quantum dot on Alice's side in concentration process of the form of $\alpha|\uparrow\rangle+\beta|\downarrow\rangle$. In this ECP,
we only need to prepare the single photon of the form of Eq. (\ref{singlephoton}). In a practical operation, it is much easier to prepare such single optical
qubit. Compared with the ECPs with linear optics \cite{zhao1,Yamamoto1}, only Alice needs to operate the whole steps and this protocol can be repeated to obtain a higher success probability.
 In our protocol, only one of
parities say Alice needs to operate the whole processing. Bob only needs to retain or discard his  particles
according to the Alice's measurement results.

In summary, we have proposed an optimal ECP with single charged
quantum dot inside an optical cavity. It has several advantages:
First, they do not need the collective measurement. Second, only one pair of less-entangled state
is required.  Third, it can be repeated to obtain a
higher success probability. Fourth, less operations and classical
communications are required . All these advantages may make this ECP
useful in current quantum information processing.

\section{Acknowledgements}
 This work is supported by the National Natural Science Foundation of
China under Grant No. 11104159,  Open Research
Fund Program of the State Key Laboratory of
Low-Dimensional Quantum Physics Scientific, Tsinghua University,
Open Research Fund Program of National Laboratory of Solid State Microstructures under Grant No. M25020 and M25022, Nanjing University,
Scientific Research
Foundation of Nanjing University of Posts and Telecommunications
under Grant No. NY211008, 210072,  University Natural Science Research
Foundation of JiangSu Province under Grant No. 11KJA510002, 11KJB510016, and the
open research fund of Key Lab of Broadband Wireless Communication
and Sensor Network Technology (Nanjing University of Posts and
Telecommunications), Ministry of Education, China, and the Project
Funded by the Priority Academic Program Development of Jiangsu
Higher Education Institutions.

\end{document}